\documentclass[11pt]{article}

\usepackage{amsmath}
\usepackage{amsfonts}

\makeatletter

\renewcommand\maketitle{\par
  \begingroup
    \if@twocolumn
      \ifnum \col@number=\@ne
        \@maketitle
      \else
        \twocolumn[\@maketitle]%
      \fi
    \else
      \newpage
      \global\@topnum\z@   
      \@maketitle
    \fi
    \thispagestyle{plain}\@thanks
  \endgroup
  \setcounter{footnote}{0}%
  \global\let\thanks\relax
  \global\let\maketitle\relax
  \global\let\@maketitle\relax
  \global\let\@thanks\@empty
  \global\let\@author\@empty
  \global\let\@date\@empty
  \global\let\@title\@empty
  \global\let\title\relax
  \global\let\author\relax
  \global\let\date\relax
  \global\let\and\relax
}

\makeatother

\newcommand{\be}{\begin{equation}}
\newcommand{\ee}{\end{equation}}

\newcommand{\n}{\ensuremath{~^{n}}}
\renewcommand\[{[\,}                            
\renewcommand\]{\,]}                            
\newcommand{\bra}[1]{\ensuremath{\langle{#1}|}}
\newcommand{\ket}[1]{\ensuremath{|{#1}\rangle}}
\newcommand{\braket}[2]{\ensuremath{\langle{#1}\mid{#2}\rangle}}
\newcommand{\V}{{\cal V}}

\let\ssection=\section
\renewcommand{\section}{\setcounter{equation}{0}\ssection}

\begin{document}

\begin{titlepage}

\title{\LARGE Histories Electromagnetism}

\author{
   Aidan Burch\footnote{Email: aidan.burch@ic.ac.uk} \\[0.5cm]
   {\normalsize The Blackett Laboratory, Imperial College, London SW7 2BZ, UK}}

\date{28.11.03}

\maketitle

\begin{abstract}\normalsize
Working within the HPO (History Projection Operator) Consistent
Histories formalism, we follow the work of Savvidou on (scalar)
field theory \cite{sav01a} and that of Savvidou and Anastopolous
on (first-class) constrained systems \cite{sav01c} to write a
histories theory (both classical and quantum) of Electromagnetism.
We focus particularly on the foliation-dependence of the histories
phase space/Hilbert space and the action thereon of the two
Poincar\'{e} groups that arise in histories field theory. We
quantise in the spirit of the Dirac scheme for constrained
systems.

\end{abstract}

\thispagestyle{empty}

\end{titlepage}


\newpage

\section{Introduction}\label{intro}

The aim of this paper is to demonstrate the application of certain
ideas and techniques that have been developed within the HPO
(History Projection Operator) histories formalism over recent
years to the theory of Electromagnetism. Specifically, we follow
up on two pieces of work which are naturally combined therein:

\textbf{Field Theory.} In \cite{sav01a}, Savvidou describes the
histories theory of the (classical and quantum) scalar field. This
has the important feature that there exist two distinct
Poincar\'{e} groups. The `internal' group is simply the histories
analogue of that of the standard theory, but there also exists an
`external' group that explicitly performs changes of the
foliation. This is important as it provides a way of relating
quantities that are defined with respect to different foliations.
These groups arise as a consequence of one of the most powerful
and interesting features of histories theories, namely that there
exist two distinct types of time transformation each of which
represents a distinct \emph{quality} of time: (a) the internal
time (`time of becoming'), which is related to the dynamics of the
particular system in question, and (b) the external time (`time of
being'), which is related to the causal ordering of events, ie.
the kinematics.\footnote{For a detailed exposition of the HPO
continuous time histories programme, the reader is referred to
\cite{sav99b}}

\textbf{Constrained Systems.} In \cite{sav01c}, Savvidou and
Anastopolous describe an algorithm for working with systems with
first-class constraints within the HPO formalism. They focussed
specifically on parameterised systems, ie. those systems whose
Hamiltonian is itself a first-class constraint, as a natural
precursor to understanding `histories' general
relativity\footnote{For progress with this enterprise, see
\cite{sav01b} \cite{sav03a} \cite{sav03b}.}, and demonstrated that
the histories on the reduced phase space \emph{retained} their
intrinsic temporal ordering. The quantisation algorithm is in the
spirit of the Dirac scheme for constrained systems.

The theory of Electromagnetism, as a field theory with first-class
constraints, thus perfectly combines the above pieces of work, but
also brings something new to each when studied within the
histories framework. In the first instance we shall see explicitly
how the histories phase space and reduced phase space depend on
the foliation and discuss the importance of the external boost in
this respect. Secondly, we will have to deal with the fact (not
tackled in detail in \cite{sav01c}) that our constraints have
continuous spectra, and thus the physical Hilbert space cannot be
a true linear subspace of the full (unconstrained) Hilbert space.

The outline of the paper is as follows: in Section 2 we give a
brief account of those aspects of the Histories programme most
relevant to our needs, and then we present the histories theory of
Electromagnetism, starting with the classical theory in Section 3,
and then it's quantisation in Section 4. We conclude in Section 5.

Finally we note that the classical history theory of vector fields
has been studied by Noltingk \cite{nol01a} as well as their BRST
quantisation \cite{nol01b}, though he follows a fundamentally
different approach which centers on defining five component vector
fields to incorporate the two times.

\section{The Histories Program}\label{hp}

The consistent histories version of quantum mechanics was
originally developed in the 80's by Griffiths \cite{gri84} and
then built on (each with different emphases) by Omnes \cite{omn88}
and then Gell-Mann and Hartle \cite{gmh90} and Hartle
\cite{gmh93}. The main aim (particularly of the latter) was to
develop a quantum mechanics of closed systems.

As formulated by Gell-Mann and Hartle, a history, $\alpha$, is
represented by a class operator, $C_{\alpha}$, that is a product
of Heisenberg picture projection operators. Dynamic information is
contained in the decoherence functional, defined on a pair of
histories as: \be
d(\alpha,\beta)=tr(C_{\alpha}^{\dag}\rho_{0}C_{\beta})\ee where
$\rho_{0}$ is the density matrix describing the initial state of
the system. If a history is part of a `consistent' set, then
probabilities (in the usual Kolmogorov sense) may be assigned to
the individual histories according to
$p(\alpha)=d(\alpha,\alpha)$.

The HPO formalism was developed initially by Isham \cite{ish94}
and Isham and Linden \cite{il94} who sought a histories version of
single-time quantum logic. To this end they re-defined the class
operator as a \emph{tensor} product of \emph{Schr\"{o}dinger}
picture operators, so it would be a genuine projection operator on
some suitable `history' Hilbert space. This formalism was extended
to the case of continuous time histories by Isham and co-workers
\cite{il95} \cite{ilss98}, in which the `history group' -
analogous to the usual Heisenberg-Weyl group - was introduced.
However, this structure lacked any clear notion of time evolution.
It was only with Savvidou's introduction of the action operator -
the quantum analogue of the classical Hamilton-Jacobi action
functional - that the temporal structure of histories theory was
established in the form as it is used now (see \cite{sav99a}). It
is these two - the history group and action operator - that are
the key elements of any history theory.

\subsection{The History Group}

By introducing the history group, an HPO theory may be seen as
seeking a suitable representation of a certain algebra, eg. for a
(non-relativistic) particle moving on the real line (see
\cite{sav99b}) and a continuous time label, $t\in\mathbb{R}$, the
(non-zero) commutation relation  is\footnote{$\hbar=1$}: \be
\[\, x_t\,, p_{t'} \,\] = i \delta(t-t')\ee This algebra is
isomorphic to that of a field theory in one spatial dimension, and
field-theoretic techniques are usefully employed to find a
suitable representation. Following Araki \cite{ara}, the proper
representation of this algebra is selected by requiring that the
Hamiltonian exist as a self-adjoint operator, and it will come as
no surprise that a Fock representation provides the necessary
`history' Hilbert space.

In histories quantum scalar field theory (see \cite{sav01a}),
after foliating Minkowski space with a unit timelike vector,
$n_{\mu}$\footnote{We use the metric signature $(+,-,-,-)$.}, we
have: \be\label{phipiscalar} \[\,\, \phi(X)\,, \pi(X') \,\,\]
=i\delta^4(X-X')\ee where we are using a `pseudo-covariant'
notation $X=n\cdot t+x_n$ ($x_n$ is a four vector such that
$n\cdot x_{n}=0$). A representation of this algebra is found in
terms of creation and annihilation, $b^{\dag}(X)$ and $b(X)$, on
the (history) Fock space,
$\V_{scalar}=exp(L^2(\mathbb{R}^4,d^4X))$. Indeed, it is found
that all foliation dependent representations exist \emph{on the
same} Fock space.

\subsection{The Action Operator}

The other key element of a histories theory is the action
operator, $S(\gamma)$ (see \cite{sav99a}). It is this that is the
generator of time transformations of an HPO theory, combining the
Liouville operator, $V(\gamma)$, which generates time translations
in the external (kinematical) time label, and the Hamiltonian,
$H(\gamma), $which generates time translations in the internal
(dynamical) time label. For our non-relativistic particle, the
action would be written: \be S(\gamma)=V(\gamma)-H(\gamma)=\int
dt~[p_t \dot{x}_t-H_t(p_t,x_t)](\gamma)\ee where $H_t$ is a one
parameter family of Hamiltonians.

In the field theory case, we have two Poincar\'{e} groups, with
the Hamiltonian being the time translations generator of the
internal group, and the Liouville being the time translations
generator of the external group. The generators of the internal
group are time-averaged versions of the generators of the standard
group, but it is the external group that is the novel object, as
the boost of this group generates \emph{changes of the foliation}
as well. Its action on the foliation-dependent scalar field is
given by: \be
^{ext}U(\Lambda)\,^n\phi(X)^{ext}U(\Lambda)^{-1}=\,^{\Lambda
n}\phi(\Lambda^{-1} X)\ee (where $U(\Lambda)$ is the unitary
operator that generates the Lorentz transformation) and thus we
have a way of relating quantities defined with respect to
different foliations.

Finally, we note, without going into great detail, that there is
an analogous formalism for classical histories\footnote{see
Chapter 5. of \cite{sav99b} for further details} which we will use
to write the classical history theory of EM below. This involves
thinking of a history as a map from the real line into the
classical phase space, $\Gamma$. A natural symplectic structure
can be defined on $\Pi$ (the history phase space), giving rise to
the Poisson algebra. The equations of motion can be expressed by
saying that, for any function $F$ on $\Pi$, their solutions,
$\gamma_{cl}$, will satisfy: \be \{F,S\}(\gamma_{cl})~=~0\ee

\subsection{The Constrained Systems Algorithm}\label{csa}

The theory of constrained systems was extensively studied by Dirac
\cite{dir64}, though we primarily use \cite{wipf} and \cite{ht92}.
In essence, a first class constraint, $\phi(x,p)=0$, is to be seen
as a generator  of gauge transformations which partitions the
phase space, $\Gamma$, (and, thus, the constraint surface) into
orbits. The reduced phase space, $\Gamma_{red}$, is then
isomorphic to the space of orbits. There exists a unique
`reduction' of a function $F$ on $\Gamma$ to a function
$\tilde{F}$ on $\Gamma_{red}$ if $F$ has a weakly vanishing
Poisson bracket with the constraint. Dirac quantisation proceeds
by constructing the unconstrained Hilbert space, $\mathcal{H}$,
writing the constraint as an operator, and then defining the
physical Hilbert space as that linear subspace (modulo
considerations of the constraint spectrum) of $\mathcal{H}$ which
is spanned by those eigenvectors of the constraint whose
corresponding eigenvalue is zero.

In histories theory \cite{sav01c} we write the  (time-averaged)
constraint as $\Phi_{\lambda}(\gamma)=\int dt
\lambda(t)\phi(x_{t},p_{t})(\gamma)$. As above, the action of the
constraint will partition $\Pi$ (the history phase space) into
orbits, and we can define $\Pi_{red}$ (the reduced phase space) as
the space of equivalence classes of histories on the constraint
surface, $C_{h}$. (Histories will be equivalent if they lie on the
same orbit). Again, there will be a unique `reduction' of any
function, $F$, on $\Pi$ to a function, $\tilde{F}$, on $\Pi_{red}$
if $\{F,\Phi_{\lambda}\}\approx 0$.

The quantisation algorithm for a histories theory follows the
spirit of the Dirac scheme, briefly described above. It is
implemented, once the constraint is suitably defined as an
operator, by first observing that we require: \be
d(e^{i\Phi_{\lambda}}\alpha
e^{-i\Phi_{\lambda}},e^{i\Phi_{\lambda}}\beta
e^{-i\Phi_{\lambda}})=d(\alpha,\beta)\ee To meet this requirement
(modulo, as above, issues concerning the nature of the constraint
spectrum) we define a projector, $E$, onto the closed linear
subspace of the (unconstrained) history space, $\V$, corresponding
to the zero eigenvalue of $\Phi_{\lambda}$ and then substitute
$\alpha$ for $E\alpha E$ in the expression for the decoherence
functional\footnote{Evidently $e^{i\Phi_{\lambda}}E=E$}.

We are now in a position to put these ideas into practice, writing
the classical histories theory of Electromagnetism in the next
section, and its quantisation in the subsequent one.

\section{Electromagnetism - Classical}

We begin with a brief review of the standard theory.

\subsection{Basics}

The EM Action is \be S=-\frac{1}{4}\int d^{4}x~
F_{\mu\nu}F^{\mu\nu} \ee where
$F_{\mu\nu}=\partial_{\mu}A_{\nu}-\partial_{\nu}A_{\mu}$. The
equations of motion are computed from setting $\delta S=0$, and we
get: \be
\partial ^{\mu}F_{\mu\nu}=0
\ee To write this in Hamiltonian form we first define the momentum
conjugate to the vector potential: \be \pi^{\mu}=\frac{\delta
L}{\delta\dot{A}_{\mu}}=F^{\mu 0} \ee So
$\pi^{i}=\partial^{i}A^{0}-\partial^{0}A^{i}=-[(\nabla
A^{0})^i+\dot{A}^i]=(\underline{E})_i$, and we have the following
constraints: \be\label{constraint1} \pi^{0}=0 \ee (this is a
primary, first-class constraint), and the Gauss Law constraint
(this is a secondary, first-class constraint),
\be\label{constraint2}
\partial_i\pi^i=0 \ee derived from the consistency condition that the
primary constraint be conserved in time, ie. $\{H,\pi^{0}\}=0$
(where $H$ is the canonical Hamiltonian given below). In terms of
the observable fields, $\underline{E}$ and
$\underline{B}=\nabla\times\underline{A}$ the canonical
Hamiltonian is written: :\be H=\int d^{3}x~
\left(\frac{1}{2}\underline{E}^{2}+\frac{1}{2}\underline{B}^{2}-A^{0}\nabla\cdot\underline{E}\right)
\ee

\subsubsection{Poincar\'{e} Invariance in the Standard Theory}

The generators of the Poincar\'{e} group of the standard theory
are `taken over' (in time-averaged form) to the histories theory
as the `internal' group. Following \cite{itz} these are derived
from the energy-momentum tensor (to which a total divergence has
to be added):
\be\tilde{\Theta}^{\mu\nu}=-F^{\mu\rho}\partial^{\nu}A_{\rho}+\frac{1}{4}\eta^{\mu\nu}F^2+\partial_{\rho}(F^{\mu\rho}A^{\nu})\ee
The ten generators of the Poincar\'{e} group are then written:
\begin{eqnarray} P^{\alpha}&=&\int d^3
x ~\tilde{\Theta}^{0\alpha}\\
M^{\alpha\beta}&=&\int d^3 x~
(x^{\alpha}\tilde{\Theta}^{0\beta}-x^{\beta}\tilde{\Theta}^{0\alpha})
\end{eqnarray}

From these, we deduce the explicit, canonical form of the
Hamiltonian, linear momentum, angular momentum and boost
generators to be, respectively:
\begin{eqnarray}
H&=&\int d^{3}x~
\left(-\frac{1}{2}\pi_i\pi^i+\frac{1}{2}(\nabla\times
\underline{A})^{2}-A^{0}\partial_i\pi^i\right)\label{canH} \\
P_i&=&\int d^3 x ~\pi^j \partial_i A_j\label{canP} \\
J^i&=&\epsilon^{ijk}\int d^3 x~(\pi^l x_j \partial_k A_l +\pi_j A_k)\label{canJ} \\
K_i&=&\int d^3 x
-x_i\left(-\frac{1}{2}\pi_j\pi^j+\frac{1}{2}(\nabla\times
\underline{A})^{2}-A^{0}\partial_j\pi^j\right)\label{canK}
\end{eqnarray}
(where we have chosen $x^0=0$ in the expression for the boost).
This algebra closes only weakly, ie. subject to the Gauss Law
constraint, a fact that will be of significance when we come to
the quantisation.

We now turn to the histories formulation of classical
electromagnetism.

\subsection{The Histories Phase Space, $\Pi$} The phase space of canonical EM is
$\Gamma=(A_i(\underline{x}),\pi_j(\underline{x}))$, and a history
is defined to be a path: \be
\gamma:\mathbb{R}\rightarrow\Gamma\ee\be
t\mapsto(A_i(t,\underline{x}),\pi_j(t,\underline{x}))\ee The space
of histories, $\Pi$, is defined to be the space of all such smooth
paths $\gamma$. As explained before we can use a `covariant-like'
notation, writing $X=n\cdot t+x_{n}$ where $n_{\mu}$ is a unit
time-like vector, so $t=n\cdot X$ and $x_{n}$ is `n-spatial', ie.
$n\cdot x_{n}=0$.

However, we wish to find a representation of the time-averaged
canonical expressions on the phase space coordinatised by
$(A_{\mu}(X),\pi^{\nu}(X))$. This is so we can write a
representation of both the internal \emph{and} external groups on
the same space. This is achieved using the $n$-spatial projector,
$P_{\mu\nu}$, introduced earlier, along with the foliating
timelike vector, $n_{\mu}$. In this notation, we write the
foliation dependent (canonical) fields $\,^n
A_{\mu}:=P^{\nu}_{\mu}A_{\nu}$ (and likewise for the conjugate
momenta. The Hamiltonian is written:\be\label{classH} H_{n}=\int
d^{4}X~
\left(\frac{1}{2}[P^{\mu\nu}\pi_{\mu}\pi_{\nu}+(\nabla^{\mu}_{\sigma}A^{\sigma})(\nabla_{\mu\delta}A^{\delta})]+n^{\rho}A_{\rho}P^{\mu\nu}\partial_{\mu}\pi_{\nu}\right)\chi(n\cdot
X)\ee (where the subscript `n' refers to the particular foliation,
and we have introduced the notation
$\nabla_{\mu\sigma}A^{\sigma}\equiv\epsilon_{\mu\nu\rho\sigma}n^{\nu}\partial^{\rho}A^{\sigma}$).
This is the generator of time translations of the internal group.
We then define the Liouville operator (the generator of time
translations in the external group): \be V_{n}=\int d^{4}X~
\pi^{\mu}n^{\rho}\partial_{\rho}A_{\mu} \ee and thus can write the
action functional: \be S_{n}=V_{n}-H_{n} \ee It is the action
functional that is to be understood as the `true' generator of
time translations of the theory, naturally intertwining the two
modes of time represented by the Hamiltonian and Liouville
operators. The fundamental Poisson brackets are now:
\be\{A_{\mu}(X),A_{\nu}(Y)\}=0=\{\pi^{\mu}(X),\pi^{\nu}(Y)\}\ee
and
\be\{A_{\mu}(X),\pi^{\nu}(Y)\}=\delta^{\nu}_{\mu}\delta^{4}(X-Y)\ee
We now turn to the central issue of Poincar\'{e} invariance.

\subsection{The Poincar\'{e} Groups}

As was the case for the scalar field, we seek representations for
two Poincar\'{e} groups on the history space, one associated with
the internal time label, and one associated with the external time
label. The generators for spatial translations and spatial
rotations will be the same for each group, so we focus our
attentions on the time translation and boost generators in each
case.

\subsubsection{The Internal Poincar\'{e} Group}

The generators of the internal Poincar\'{e} group will be
time-averaged versions of the generators of the standard theory
(Eqs.~\ref{canH}-\ref{canK}). The time translation generator is,
of course, just the Hamiltonian of Eq.~\ref{classH} and we define
the boost at $s=0$ as: \be\label{classintK}\begin{split}
^{int}K(m)=-m_{\mu}\int d^{4}X~
X^{\mu}(\frac{1}{2}[P^{\mu\nu}\pi_{\mu}\pi_{\nu}+(\nabla^{\mu}_{\sigma}A^{\sigma})(\nabla_{\mu\delta}A^{\delta})]+
\\n^{\rho}A_{\rho}P^{\mu\nu}\partial_{\mu}\pi_{\nu})\end{split}\ee
where $m_{\mu}$ is a space-like vector, ie. $n\cdot m=0$,
parameterising the boost. Given a function $A$ on $\Pi$, we can
denote the one parameter group of transformations it generates as
$s\mapsto T_A(s)$ and its action on the algebra of functions, $B$,
as: \be
T_A(s)[B]=\sum_{n}\frac{s^n}{n!}\underbrace{\{A,\{A,\ldots\{A,}_{n-times}B\}\ldots\}\}\ee
So, we first define the classical analogue of the Heisenberg
picture fields by: \be T_H(s)[\,^n A_{\mu}(X)]=\,^n
A_{\mu}(X,s)\ee\be T_H(s)[\,^n\pi^{\mu}(X)]=\,^n\pi^{\mu}(X,s)\ee
and can now see explicitly the sense in which the Hamiltonian
generates time translations in the internal time label by looking
at its action on the `Heisenberg' picture fields: \be
T_H(\tau)[\,^n A_{\mu}(X,s)]=\,^n A_{\mu}(X,s+\tau)\ee \be
T_H(\tau)[\,^n\pi^{\mu}(X,s)]=\,^n\pi^{\mu}(X,s+\tau)\ee

The internal boost generator will mix the internal time
parameter,`$s$', with the spatial coordinates:
\begin{eqnarray}T_{^{int}K(m)}[\,^n A_{\mu}(X,s)]&=&\,^n A_{\mu}(\Lambda^{-1}(X,s))
\\
T_{^{int}K(m)}[\,^n\pi^{\mu}(X,s)]&=&\,^n\pi^{\mu}(\Lambda^{-1}(X,s))\end{eqnarray}
where $\Lambda^{-1}(X,s)$ is related to $(X,s)$ (the time label
`$t$' is, of course, constant) by the Lorentz boost parameterised
by $m^{\mu}$, ie. the velocity of the moving frame is given by:
\be v^i=c\frac{tanh|m|m^i}{|m|}\ee

\subsubsection{The External Poincar\'{e} Group}

In contrast to the definition of the generators of the internal
group, we use the covariant fields, $(A_{\mu}(X),\pi^{\mu}(X))$ in
the definition of the generators for the external Poincar\'{e}
group. These are: \be P^{\mu}=\int
d^{4}X~\pi^{\nu}\partial^{\mu}A_{\nu}\ee and \be M^{\mu\nu}=\int
d^{4}X~[\pi^{\rho}(X^{\mu}\partial^{\nu}-X^{\nu}\partial^{\mu})A_{\rho}]+\sigma^{\mu\nu}\ee
where $\sigma^{\mu\nu}$ is the spin term, given by:
\be\sigma^{\mu\nu}=\int
d^4X~(\pi^{\mu}A^{\nu}-\pi^{\nu}A^{\mu})\ee As before, we are
particularly interested in the actions of the time translation
generator $V=P^{0}$ and the boosts generator
$K(m)=n_{\mu}m_{\nu}M^{\mu\nu}$. These are therefore written: \be
V=\int d^{4}X~\pi^{\mu}n^{\nu}\partial_{\nu}A_{\mu}\ee and \be
^{ext}K(m)=m_{\mu}\int d^{4}X~[(n\cdot
X)\pi^{\nu}\partial^{\mu}A_{\nu}-X^{\mu}\pi^{\rho}n^{\nu}\partial_{\nu}A_{\rho}]+n_{\mu}m_{\nu}\sigma^{\mu\nu}\ee
The effect of the Liouville functional is to generate the
following algebra automorphisms, in which we can clearly see that
it generates time translation in the external time label:
 \be T_V(\tau)[A_{\mu}(X,s)]=e^{-\tau
n_{\sigma}\partial^{\sigma}}A_{\mu}(X,s)=A_{\mu}(X',s)\ee\be
T_V(\tau)[\pi^{\mu}(X,s)]=e^{-\tau
n_{\sigma}\partial^{\sigma}}\pi^{\mu}(X,s)=\pi^{\mu}(X',s)\ee
where $X'$ is the point in $\mathcal{M}$ associated with the pair
$(\underline{x},t+\tau)$.

Let us now turn to the transformations generated by the external
boosts. These will mix the external time parameter, `$t$', with
the spatial coordinates. The finite transformations can be
written:\be\label{extKA}
T_{^{ext}K(m)}[A_{\mu}(X,s)]=\Lambda_{\mu}^{\nu}A_{\nu}(\Lambda^{-1}(X),s)\ee
\be\label{extKpi}
T_{^{ext}K(m)}[\pi^{\mu}(X,s)]=\Lambda^{\mu}_{\nu}\pi^{\nu}(\Lambda^{-1}(X),s)\ee

As previously stated, the role of the external group is an
interesting one, and it is this that is one of the novel features
of histories field theory. The effect of the external boosts is to
mix the spatial coordinate with the external time label `$t$' and,
as the phase space has an implicit foliation dependence, it will
also boost the foliation vector itself, thus generating
transformations between different foliation-dependent
representations.

\subsection{The Reduced Phase Space, $\Pi_{red}$}

Our next task is to follow the algorithm of \cite{sav01c} to
ascertain a suitable description of the reduced phase space,
$\Pi_{red}$ on which the true degrees of freedom of the theory are
defined. To this end, we are interested in the actions of the
constraints on the phase space (and in particular the history
constraint surface, $C_{h}$) because, by examining their action,
we can define suitable coordinates (ie. ones constant along the
orbits) for the reduced phase space $\Pi_{red}$.

We write the time-averaged analogues of the constraints of the
standard theory as follows: \be \Psi_{\lambda}=\int
d^{4}X~\lambda(X)n_{\mu}\pi^{\mu}\approx0\ee\be
\Phi_{\lambda}=\int
d^{4}X~\lambda(X)P^{\mu\nu}\partial_{\mu}\pi_{\nu}\approx0\ee and
consider their action on the coordinates of $\Pi$. Under
$\Psi_{\lambda}$ we have: \be\label{actPsi}
(A_\mu(X),\pi^{\mu}(X))\rightarrow(A_\mu(X)-\lambda(X)n_{\mu},\pi^{\mu}(X))\ee
Under $\Phi_{\lambda}$ we have:
\be\label{actPhi}(A_\mu(X),\pi^{\mu}(X))\rightarrow(A_\mu(X)+P^{\rho}_{\mu}\partial_{\rho}\lambda(X),\pi^{\mu}(X))\ee
Evidently $\pi^{\mu}(X)$ is constant along the orbits, so we just
seek a quantity associated with the vector potential that is gauge
invariant.

Eqs.~\ref{actPsi} and \ref{actPhi} tell us that the transverse
components of the vector potential remain constant along the
orbits of the constraints and are thus good coordinates for
$\Pi_{red}$, whereas the scalar and longitudinal components
correspond to the degenerate directions of $\Psi_{\lambda}$ and
$\Phi_{\lambda}$ respectively. (This state of affairs is more
clearly seen if we use a Fourier transform and work in momentum
space). If we combine this knowledge with a look at the
constraints themselves, which (if we were to Fourier transform
them) readily show us that the constraint surface, $C_h$, is
defined by $\pi^0=\pi^L=0$, where these are respectively the
scalar and longitudinal components of the conjugate momentum, we
can deduce that $\Pi_{red}$ is suitably coordinatised by
$(A^{\perp}_{\mu}(X),\pi^{\perp}_{\mu}(X))$, where the superscript
`$\perp$' indicates the transverse components, and these are
defined by: \be
A^{\perp}_{\mu}(X)=\left(\frac{^n\partial_{\mu}\,^n\partial^{\nu}}{^n\Delta}-P^{\nu}_{\mu}\right)A_{\nu}(X)\ee
(and likewise for $\pi^{\perp}$) and where $^n\partial_{\mu}$ is
shorthand for $P_{\mu}^{\alpha}\partial_{\alpha}$ and the
(invertible) partial differential operator $^n\Delta$ is defined:
\be\label{defDelta} (\,^n\Delta
f_{\rho})(X)=(P^{\mu\nu}\partial_{\mu}\partial_{\nu})f_{\rho}(X)\ee

The (non-zero) Poisson bracket relation on the reduced phase space
is given by: \be
\{A^{\perp}_{\mu}(X),\pi^{\perp\nu}(X')\}=T^{\nu}_{\mu}\delta^4(X-X')\ee
where: \be
(T^{\nu}_{\mu}f_{\nu})(X)\equiv\left(\frac{^n\partial_{\mu}\,^n\partial^{\nu}}{^n\Delta}-P^{\nu}_{\mu}\right)f_{\mu}(X)\ee

We are now in a position to examine whether or not we can write a
representation of the two Poincar\'{e} groups on $\Pi_{red}$.

\subsection{The Reduced Poincar\'{e} Algebras}

As explained in Section \ref{csa}, for a function on the whole
phase space to reduce to a corresponding function on the reduced
phase space, it is necessary that its Poisson bracket with the
constraints is weakly zero. We expect to find that the generators
of the \emph{internal} Poincar\'{e} group reduce to $\Pi_{red}$.
However, we do not expect to find a full representation of the
\emph{external} Poincar\'{e} group on the reduced phase space. In
\cite{sav01a} the foliation dependence of the phase space was
emphasised but not explicit. In the case of EM we shall see this
dependence explicitly as the action of the external boost will
affect the definition of $\Pi_{red}$ and so we do not expect to
find a reduced version of this generator. We now turn to the
explicit results.

As before, we are only interested in the time translation and
boost generators of each Poincar\'{e} group and thus we need only
compute the Poisson brackets of $S,~^{int}K(m)$ and $^{ext}K(m)$
with the constraints. We find the following results (recall that
$\Psi_{\lambda}$ is the `$\pi^{0}$' constraint and
$\Phi_{\lambda}$ the Gauss Law constraint): \be
\{S,\Psi_{\lambda}\}=\Psi_{\dot{\lambda}}-\Phi_{\lambda}\approx0\ee
and \be \{S,\Phi_{\lambda}\}=\Phi_{\dot{\lambda}}\approx0\ee So
the action functional weakly commutes with both constraints and so
can be reduced to a functional $\tilde{S}$ acting on $\Pi_{red}$.

The internal boost generator has the following Poisson brackets
with the constraints: \be
\{^{int}K(m),\Psi_{\lambda}\}=\Phi_{-m_{\alpha}X^{\alpha}\lambda}\approx0\ee
and \be \{^{int}K(m),\Phi_{\lambda}\}=0\ee This in line with what
we expected, ie. that the generators of the internal Poincar\'{e}
group commute with the constraints and thus we have a
representation of the internal group on $\Pi_{red}$. (Of course,
something would be quite amiss if we did not have this as the
internal group is the histories analogue of the Poincar\'{e} group
of standard Maxwell theory).

The external boost generator forms the following Poisson brackets
with the constraints: \be
\{^{ext}K(m),\Psi_{\lambda}\}=\Psi_{(n_{\beta}X^{\beta}m_{\alpha}\partial^{\alpha}-m_{\beta}X^{\beta}n_{\alpha}\partial^{\alpha})\lambda}
-\int d^{4}X~\lambda(X)m_{\alpha}\pi^{\alpha}\ee and \be
\{^{ext}K(m),\Phi_{\lambda}\}=\Psi_{m_{\alpha}\partial^{\alpha}\lambda}+\int
d^{4}X~(n_{\alpha}\partial^{\alpha}\lambda(X))m_{\beta}\pi^{\beta}\ee
Neither of these are weakly zero, and so the external boost
generator cannot be reduced to $\Pi_{red}$.

For those functions that \emph{can} be reduced, we use the
coordinates for the reduced phase space that we worked out in the
previous section. The Hamiltonian and Liouville functionals on the
reduced phase space are written as follows:
\begin{eqnarray}\tilde{H}&=&\int
d^4X~\frac{1}{2}\left(\pi^{\perp\mu}\pi^{\perp}_{\mu}+A_{\mu}^{\perp}\,^n\Delta
A^{\perp\mu}\right)\label{redH}\\ \tilde{V}&=&\int
d^4X~\pi^{\perp\mu}n_{\nu}\partial^{\nu}A_{\mu}^{\perp}\label{redV}\end{eqnarray}where
we have used, in the expression for the Hamiltonian: \be
A_{\mu}\,^n\Gamma^{\mu\nu}A_{\nu}=A^{\perp}_{\mu}\,^n\Delta
A^{\perp\mu}\ee with $^n\Delta$ defined in Eq.~\ref{defDelta}.
Thus the action functional on $\Pi_{red}$ is written:
\be\tilde{S}~=~\tilde{V}~-~\tilde{H}\ee and the classical paths
which are solutions to the equations of motion are those which
satisfy: \be \{\tilde{S},\tilde{F}\}(\gamma_{cl})~=~0\ee for all
functions $\tilde{F}$ defined on $\Pi_{red}$.

\section{Electromagnetism - Quantisation}\label{quantintro}

For the quantisation of the theory we continue to follow the
algorithm laid down by Savvidou and Anastopolous, which, as
outlined in Section \ref{csa}, essentially follows the Dirac
scheme. We define the history space, $\V$, by consideration of the
history group, and define the constraints thereon. However, as we
mentioned, the constraints have continuous spectra, and thus the
physical Hilbert space, $\V_{phys}$, will not be a genuine
subspace of the history Hilbert space. This will be explicitly
demonstrated. And so we are lead to a creative implementation of
the algorithm\footnote{The central idea here is due to Savvidou -
private communication.}, in which the physical Hilbert space is
defined separately, based on an analysis, in terms of coherent
states, of how the constraints act on $\V$. Appropriate mappings
are then defined between $\V$ and $\V_{phys}$ such that objects on
one can be related to objects on the other.

\subsection{The History Hilbert Space $\V$}\label{V}

So the first stage is to define the History Hilbert space.
Following the methods of \cite{il95} and \cite{sav01a}, we start
by defining the History Algebra:
\begin{eqnarray}
[\, A_{\mu}(X)\,, A_{\nu}(X')\,] &=& 0    \\
\[\, \pi_{\mu}(X)\,, \pi_{\nu}(X') \,\] &=& 0   \\
\[\, A_{\mu}(X)\,, \pi^{\nu}(X') \,\] &=& i \delta^{\nu}_{\mu}\delta^4(X-X')
\end{eqnarray}
or, in its more rigorous, smeared form:
\begin{eqnarray}
[\, A_{\mu}(f^{\mu})\,, A_{\nu}(f'^{\nu})\,] &=& 0  \label{haAA}  \\
\[\, \pi_{\mu}(h^{\mu})\,, \pi_{\nu}(h'^{\nu}) \,\] &=& 0  \label{hapipi} \\
\[\, A_{\mu}(f^{\mu})\,, \pi^{\nu}(h_{\nu})\,\] &=& i
\int~d^4X\delta^{\nu}_{\mu}f^{\mu}(X)h_{\nu}(X) \label{haApi}
\end{eqnarray}
where $f^{\mu}(X),~h_{\mu}(X)$ are elements of a suitable space of
smearing functions which we will leave unspecified beyond saying
that it must at least be a subspace of $\oplus_{i=1\ldots
4}L^2_{\mathbb{R}}(\mathbb{R}^4,d^4X)_i$. Let us denote this space
$\mathcal{T}_{\mathbb{R}}$. It is natural to seek a Fock
representation of this algebra, and this is achieved by first
taking the complexification of the space of smearing functions,
ie.
$\mathcal{T}_{\mathbb{C}}=\mathcal{T}_{\mathbb{R}}\oplus\mathcal{T}_{\mathbb{R}}$
and then exponentiating the resulting space to give
$\V=e^{\mathcal{T}_{\mathbb{C}}}$. The Fock space thus defined
will carry a natural representation of the above History Algebra,
which we seek explicitly below, in terms of creation and
annihilation operators: \be\[\, b_{\mu}(X)\,, b^{\dag\nu}(X)
\,\]=\delta^{\nu}_{\mu}\delta^4(X-X')\ee

\subsubsection{The Representation of $(A,\pi)$ in terms of
$(b^{\dag},b)$}\label{fieldrep}

We can easily write a representation of the fully covariant
fields, $(A_{\mu}(X),\pi^{\mu}(X))$:
\begin{eqnarray}
A_{\mu}(X)&=&\frac{1}{\sqrt{2}}(b_{\mu}(X)+b_{\mu}^{\dag}(X)) \label{covAbb}\\
\pi_{\mu}(X)&=&
-\frac{i}{\sqrt{2}}(b_{\mu}(X)-b_{\mu}^{\dag}(X))\label{covpibb}
\end{eqnarray}
However, what we require in order to define the Hamiltonian is
foliation-dependent fields. So we start from a normal-ordered
analogue of the classical unconstrained Hamiltonian: \be
^nH=:\frac{1}{2}\int~d^4X \left(P^{\mu\nu}\, ^n\pi_{\mu}
\,^n\pi_{\nu}+ \,^nA_{\mu} \,^n\Gamma^{\mu\nu}\,
^nA_{\nu}\right):\ee and may think, at first, to define:
\begin{eqnarray}
^nA_{\mu}(X)&=&\frac{1}{\sqrt{2}} (\,^n\Gamma^{\nu}_{\mu})^{-\frac{1}{4}}(b_{\nu}(X)+b_{\nu}^{\dag}(X)) \\
^n\pi_{\mu}(X)&=& -\frac{i}{\sqrt{2}}
(\,^n\Gamma^{\nu}_{\mu})^{\frac{1}{4}}(b_{\nu}(X)-b_{\nu}^{\dag}(X))
\end{eqnarray}
However, there is a problem here, as the operator
$\Gamma^{\mu\nu}$ has zero eigenvalues, and is, therefore, not
invertible. To see this, it is easier to use `canonical notation',
ie.
$\Gamma^{ij}=\partial^i\partial^j-\delta^{ij}\partial_k\partial^k$.
We then examine the action of this operator on an element,
$f_i(x)$, of the smearing function space - which we split into
it's transverse and longitudinal components,
$f_i=f^{\perp}_i+f^{\parallel}_i$ - and find: \be
\Gamma^{ij}f_i(x)=\Gamma{ij}(f^{\perp}_i(x)+f^{\parallel}_i(x))=\Gamma^{ij}f^{\perp}_i(x)\ee
So the longitudinal components of the smearing functions are the
zero eigenvectors of $\Gamma^{ij}$. If we now split the
Hamiltonian into its transverse and longitudinal parts, we find
(reverting to the full `histories' notation, and dropping the $^n$
superscript for ease): \be\label{splitH}
H=\frac{1}{2}\int~d^4X\left(\pi^{\perp}_{\mu}\pi^{\perp\mu}+A^{\perp}_{\mu}\,
^n\Delta
A^{\perp\mu}+\pi^{\parallel}_{\mu}\pi^{\parallel\mu}\right)\ee
where the operator $^n\Delta$ was defined in Eq.~\ref{defDelta}.
And now we see that the transverse part of the Hamiltonian is, in
essence, that of the usual `harmonic oscillators', whereas the
longitudinal part is that of a `free particle'. This form now
prompts us towards the correct definition of the fields in terms
of the creation and annihilation operators:
\begin{eqnarray}
^nA_{\mu}(X)&=&\frac{1}{\sqrt{2}} \,^n\Delta^{-\frac{1}{4}}(b_{\mu}(X)+b_{\mu}^{\dag}(X)) \\
^n\pi_{\mu}(X)&=& -\frac{i}{\sqrt{2}}\,
^n\Delta^{\frac{1}{4}}(b_{\mu}(X)-b_{\mu}^{\dag}(X))
\end{eqnarray}

Now, on the one hand, we can consider the Fock space in terms of
the orthonormal basis obtained by continual application  of the
creation operator on a translationally invariant vacuum state
(defined by $b_{\mu}\ket {0}=0$). However, it has also proved very
useful to consider the Fock space in terms of coherent states -
indeed, in \cite{il95}, these were vital to the demonstration that
there exists a natural isomorphism between an exponential Hilbert
space and the `continuous tensor product' of Hilbert spaces so
vital to the Histories programme. In the next section, we make use
of the technology of coherent states\footnote{An excellent
reference is \cite{ks}} as we seek to define the Physical Hilbert
space.

\subsection{The Physical Hilbert Space $\V_{phys}$}\label{vphys}

Recall that the constraints are written: \begin{eqnarray}
\Psi_{\lambda}&=&\int d^{4}X~\lambda(X)n_{\mu}\pi^{\mu} \\
\Phi_{\lambda}&=&\int
d^{4}X~\lambda(X)P^{\mu\nu}\partial_{\mu}\pi_{\nu}\end{eqnarray}
Substituting for the fields in terms of the creation and
annihilation operators, these become: \begin{eqnarray}
\Psi_{\lambda}&=&\frac{-i}{2}\int d^{4}X~\lambda(X)n_{\mu}\,
^n\Delta^{\frac{1}{4}}(b^{\mu}-b^{\dag\mu}) \\
\Phi_{\lambda}&=&\frac{-i}{2}\int
d^{4}X~\lambda(X)P^{\mu\nu}\partial_{\mu}\,
^n\Delta^{\frac{1}{4}}(b_{\nu}-b_{\nu}^{\dag})\end{eqnarray} It is
clear that these constraints are self-adjoint operators on $\V$,
and also that they have continuous spectra, so the physical
Hilbert space will not be a genuine subspace. However, by
consideration of the Fock space, $\V$, in terms of coherent states
we are led naturally to the correct definition of $\V_{phys}$, and
explicitly show in what sense the latter is not a true subspace of
the former.

The Weyl operator which generates the (overcomplete) set of
coherent states is written:
\begin{eqnarray}
U[f,h]&=&exp[i(\,^nA_{\mu}(f^{\mu})-\,^n\pi_{\nu}(h^{\nu}))]\ket{0}\label{weylApi}
\\&=&exp[b^{\dag}_{\mu}(z^{\mu})-b_{\mu}(z^{\ast\mu}]\ket{0}\label{weylbb}\end{eqnarray}
with
$z_{\mu}(X)=\frac{1}{\sqrt{2}}\left(\,^n\Delta^{\frac{1}{4}}h_{\mu}(X)+i\,^n\Delta^{-\frac{1}{4}}f_{\mu}(X)\right)$.
The un-normalised coherent states on $\V$ are defined for each
$z^{\mu}(X)\in\mathcal{T}_{\mathbb{C}}$ as: \be
\ket{exp~z}=e^{b^{\dag}_{\mu}(z^{\mu})}\ket{0}\ee Their overlap is
given by: \be\braket{exp~z}{exp~z'}=e^{<z,z'>}\ee (where the inner
product is $<z,z'>=\int d^4X~z_{\mu}^{\ast}(X)z'^{\mu}(X)$) and
there exists a measure, $d\sigma[z]$, such that: \be 1~=~\int
\ket{exp~z}\bra{exp~z}~d\sigma[z]\ee (That this measure exists was
demonstrated in \cite{il95}). With the aid of this resolution of
unity, we can thus define an integral representation of $\V$ in
terms of wave functionals, $\psi[z]$: \be
\ket{\psi}=\int\ket{exp~z}\braket{exp~z}{\psi}~d\sigma[z]=\int\psi[z]\ket{exp~z}~d\sigma[z]\ee
Furthermore, we can write a differential representation of a
general operator, $O$ on $\V$: \be
\bra{exp~z}:O(b^{\dag}_{\mu},b_{\mu}):\ket{\psi}=O\left(z^{\ast}_{\mu},\frac{\delta}{\delta
z^{\ast}_{\mu}}\right)\psi[z]\ee

Given this last construction, we can now rewrite the constraint
operators as follows: \be \bra{exp~z}\Psi_{\lambda}\ket{\psi}=\int
d^4X~g_{\mu}\left(\frac{\delta}{\delta
z^{\ast\mu}}-z^{\ast\mu}\right)\psi[z] \ee (where
$g_{\mu}(X)=\frac{-i}{2}\lambda(X)\,^n\Delta^{\frac{1}{4}}n_{\mu}$)
and, similarly: \be\bra{exp~z}\Phi_{\lambda}\ket{\psi}=\int
d^4X~w_{\mu}\left(\frac{\delta}{\delta
z^{\ast\mu}}-z^{\ast\mu}\right)\psi[z]\ee (where
$w_{\mu}(X)=\frac{i}{2}P^{\nu}_{\mu}\partial_{\nu}\lambda(X)\,^n\Delta^{\frac{1}{4}}$).

So now we can consider the action of the constraints on a general
wave functional $\psi[z]$, finding: \be
e^{i\Psi_{\lambda}}\psi[z]=e^{-\frac{1}{2}<g,g>-i<z,g>}\psi[z+ig]\ee
and, similarly: \be
e^{i\Phi_{\lambda}}\psi[z]=e^{-\frac{1}{2}<w,w>-i<z,w>}\psi[z+iw]\ee
We can now explicitly see that $\V_{phys}$ will not be a subspace
of $\V$ as we require the subspace to be invariant under the
action of the constraints, and are thus essentially looking for
solutions to the pair of equations:
\begin{eqnarray}
\psi[z]&=&\psi[z+ig] \\
\psi[z]&=&\psi[z+iw] \end{eqnarray} The solutions to these will be
$\psi[z^{\perp}]$, where $z_{\mu}^{\perp}(X)$ are only the
transverse components of $z_{\mu}(x)$ defined: \be
z^{\perp}_{\mu}(X)=\left(\frac{^n\partial_{\mu}\,^n\partial^{\nu}}{^n\Delta}-P^{\nu}_{\mu}\right)z_{\nu}(X)\ee
(where $^n\partial_{\mu}$ is just shorthand for
$P^{\rho}_{\mu}\partial_{\rho}$). However, it is clear that the
corresponding wave-functionals, $\psi[z^{\perp}]$, will not be
square integrable. To see this, we need only consider:
$\int\int\int |\psi[z^{\perp}]|^2
d\sigma[z^{\perp}]d\sigma[z^0]d\sigma[z^{\parallel}]$ which will
be infinite on account of the contributions from the integrations
over the scalar and longitudinal parts. This leads us to the
conclusion that what we need to do is to construct the physical
Hilbert space \emph{separately} so that the wave-functionals,
$\psi[z^{\perp}]$, \emph{are} square-integrable, and then define a
suitable mapping from $\V$ to $\V_{phys}$.

Equipped with what we know from the classical theory, and what we
have ascertained from the analysis above, we construct $\V_{phys}$
in the usual way - firstly by positing the algebra:
\begin{eqnarray}
[\, A_{\mu}(f^{\perp\mu})\,, A_{\nu}(f'^{\perp\nu})\,] &=& 0    \\
\[\, \pi_{\mu}(h^{\perp\mu})\,, \pi_{\nu}(h'^{\perp\nu}) \,\] &=& 0   \\
\[\, A_{\mu}(f^{\perp\mu})\,, \pi^{\nu}(h^{\perp}_{\nu}) \,\] &=& i
\int~d^4X\delta^{\nu}_{\mu}f^{\perp\mu}(X)h^{\perp}_{\nu}(X)
\end{eqnarray}
where the smearing functions belong to
$\mathcal{T}^{\perp}_{\mathbb{R}}=L^2_{\mathbb{R}}(\mathbb{R}^4,d^4X)\oplus
L^2_{\mathbb{R}}(\mathbb{R}^4,d^4X)$. We then take the
complexification of this space
$\mathcal{T}^{\perp}_{\mathbb{C}}=\mathcal{T}^{\perp}_{\mathbb{R}}\oplus\mathcal{T}^{\perp}_{\mathbb{R}}$
and exponentiate the resulting space to give
$\V_{phys}=e^{\mathcal{T}^{\perp}_{\mathbb{C}}}$. We can then
write a representation of the transverse fields in terms of the
creation and annihilation operators of this Fock space:
\begin{eqnarray}
^nA^{\perp}_{\mu}(X)&=&\frac{1}{\sqrt{2}} \,^n\Delta^{-\frac{1}{4}}(b^{\perp}_{\mu}(X)+b_{\mu}^{\perp\dag}(X)) \label{TAbb} \\
^n\pi^{\perp}_{\mu}(X)&=& -\frac{i}{\sqrt{2}}\,
^n\Delta^{\frac{1}{4}}(b^{\perp}_{\mu}(X)-b_{\mu}^{\perp\dag}(X))\label{Tpibb}
\end{eqnarray}
where: \be\[\, b_{\mu}(z^{\perp\mu})\,,
b^{\dag\nu}(z^{\prime\perp}_{\nu})
\,\]=<z^{\perp},z^{\prime\perp}>\ee and
$z^{\perp}_{\mu}(X)=\frac{1}{\sqrt{2}}\left(\,^n\Delta^{\frac{1}{4}}h^{\perp}_{\mu}(X)+i\,^n\Delta^{-\frac{1}{4}}f^{\perp}_{\mu}(X)\right)$

In direct analogy to $\V$, we can consider $\V_{phys}$ in terms of
the un-normalised coherent states defined by: \be\label{Tcsdef}
\ket{exp~z^{\perp}}_{\V_{phys}}~=~e^{b^{\mu}(z^{\perp}_{\mu})}\ket{0}\ee
and these will admit a resolution of unity: \be 1~=~\int
\ket{exp~z^{\perp}}\bra{exp~z^{\perp}}~d\sigma[z^{\perp}]\ee and
thus an integral representation for $\ket{\psi}\in\V_{phys}$: \be
\ket{\psi}~=~\int~\psi[z^{\perp}]\ket{exp~z^{\perp}}_{\V_{phys}}~d\sigma[z^{\perp}]\ee

We now define a mapping between $\V$ and $\V_{phys}$:
\begin{eqnarray}
L:~\V&\longrightarrow&\V_{phys}\nonumber \\
\ket{exp~z}_{\V}&\mapsto&L(\ket{exp~z}_{\V})\equiv\ket{exp~z^{\perp}}_{\V_{phys}}\end{eqnarray}
where $\ket{exp~z^{\perp}}_{\V_{phys}}$ is defined as in
Eq.~\ref{Tcsdef}. We define the (continuous) dual mapping: \be
L^{\dag}:~\V_{phys}^*\longrightarrow\V^*\ee by: \be
_{\V_{phys}}\bra{exp~z^{\perp}}L^{\dag}\ket{exp~w}_{\V}=\,_{\V_{phys}}\bra{exp~z^{\perp}}L\ket{exp~w}_{\V}\ee
We can now use these maps (and the fact that, due to the Riesz
Lemma (see eg. \cite{rs}), there is an isomorphism between a
Hilbert space, $\mathcal{H}$, and the space of continuous linear
functionals, $\mathcal{H}^*$, from $\mathcal{H}$ to $\mathbb{C}$)
to relate objects on $\V_{phys}$ to objects on $\V$:
\be\label{bVbVphys} b_{\V_{phys}}=Lb_{\V}L^{\dag}\ee Having now
established the relationship between the `full' Hilbert space,
$\V$, and the `physical' Hilbert space, $\V_{phys}$, we can now
turn to the issue of Poincar\'{e} invariance and use
Eq.~\ref{bVbVphys} to define the action operator on $\V_{phys}$.

\subsection{The Poincar\'{e} Groups}\label{pg}

In the case of classical histories electromagnetism, we proved the
existence of the two Poincar\'{e} groups on the histories phase
space, $\Pi$, and analysed their `reduction' to the reduced phase
space, $\Pi_{red}$, by considering their compatibility with the
constraints. We demonstrated the existence of the internal group
on $\Pi$, finding that the algebra closed only weakly, ie. was
only satisfied on the constraint surface. We then proved the
existence of a `reduced' internal Poincar\'{e} group on
$\Pi_{red}$, with the generators written in terms of the
transverse components of the fields. We also demonstrated the
existence of the external Poincar\'{e} on $\Pi$, but found that
the external boost generator did not commute with the constraints,
and thus could not be represented on $\Pi_{red}$. This, as we
shall see in greater detail in the quantum case below, results
from the fact that $\Pi$ and $\Pi_{red}$ are foliation dependent,
and that the external boost boosts the foliation vector as well.
So let us now discuss the issue of Poincar\'{e} invariance in the
quantum theory.

\subsubsection{The Internal Poincar\'{e} Group}\label{intpg} Our
starting point for the internal Poincar\'{e} group is (a normal
ordered version of) the unconstrained Hamiltonian given in
Section~\ref{fieldrep} and repeated here: \be
H=\frac{1}{2}:\int~d^4X\left(\pi^{\perp}_{\mu}\pi^{\perp\mu}+A^{\perp}_{\mu}\,
^n\Delta A^{\perp\mu}+\pi^{\parallel}_{\mu}\pi^{L\mu}\right):\ee
In terms of the creation and annihilation operators
(Eqs.~\ref{TAbb}-\ref{Tpibb}), this reads: \be
H=\int~d^4X\left[b^{\perp\dag}_{\mu}\,^n\Delta^{\frac{1}{2}}b^{\perp\mu}-\frac{1}{4}\left((b^{\parallel}_{\mu}-b^{\parallel\dag}_{\mu})\,^n\Delta^{\frac{1}{2}}(b^{\parallel\mu}-b^{\parallel\dag\mu})\right)\right]\ee
However, whilst the transverse part can easily be shown to exist
in the usual way, the longitudinal part does not generate
automorphisms which are unitarily implementable on account of the
presence of terms quadratic in $b_{\mu}$ and $b^{\dag}_{\mu}$. And
so the Hamiltonian does not exists on $\V$ as a self-adjoint
operator. Of course, this is no tragedy and we half-expected it
anyway as we had already seen in the classical case that the
algebra of the internal group closed only weakly.

What is important is that a representation of the internal group
can be found on $\V_{phys}$. This is straightforward. The
generators are taken straight from the classical case, suitably
ordered and then written in terms of $b^{\perp}_{\mu}$ and
$b^{\perp\dag}_{\mu}$ using Eqs.~(\ref{TAbb}-\ref{Tpibb}). They
are:
\begin{eqnarray}
\tilde{H}&=&\int~d^4X~b^{\perp\dag}_{\mu}\,^n\Delta^{\frac{1}{2}}b^{\perp\mu}\label{HVphys} \\
\tilde{P}(m)&=&im_{\nu}\int~d^4X~b^{\perp\dag}_{\mu}\partial^{\nu}b^{\perp\mu} \\
\tilde{J}(m)&=&i\epsilon_{\mu\nu\rho\sigma}n^{\mu}m^{\nu}\int~d^4X(b^{\perp\dag}_{\alpha}X^{\rho}\partial^{\sigma}b^{\perp\alpha}+b^{\perp\dag\rho}b^{\perp\sigma})
\\
\tilde{K}(m)&=&m_{\nu}\int~d^4X~b^{\perp\dag}_{\mu}\,^n\Delta^{\frac{1}{4}}X^{\nu}\,^n\Delta^{\frac{1}{4}}b^{\perp\mu}
\\
\end{eqnarray}
where we have used an obvious shorthand for operators on
$\V_{phys}$ (see Eq.~\ref{bVbVphys}), ie. \be b^{\perp\mu}\equiv
b^{\mu}_{\V_{phys}}=Lb^{\mu}_{\V}L^{\dag}\ee The analysis of this
group is essentially the same as the classical case. We define the
Heisenberg picture fields:
\begin{eqnarray} b^{\perp}_{\mu}(X,s)&=&e^{is\tilde{H}}b^{\perp}_{\mu}(X)e^{-is\tilde{H}} \\
b^{\perp\dag}_{\mu}(X,s)&=&e^{is\tilde{H}}b^{\perp\dag}_{\mu}(X)e^{-is\tilde{H}}
\end{eqnarray}
The Hamiltonian generates transformations in the internal time
label `$s$', and the internal boost mixes the internal time
parameter with the spatial coordinates. These transformations all
happen at constant `$t$' (where `$t$' is the external time
parameter).

\subsubsection{The External Poincar\'{e} Group}\label{extpg} As in
the case of the scalar field, one of the novel features of
histories theories is the existence of a second Poincar\'{e} group
- the external group - that is associated with the external time
label, `$t$'. Again we start from the classical expressions,
suitably ordered: \begin{eqnarray} P^{\mu}&=&:\int
d^{4}X~\pi^{\nu}\partial^{\mu}A_{\nu}:\\ M^{\mu\nu}&=&:\int
d^{4}X~\left[\pi^{\rho}(X^{\mu}\partial^{\nu}-X^{\nu}\partial^{\mu})A_{\rho}+(\pi^{\mu}A^{\nu}-\pi^{\nu}A^{\mu})\right]:\end{eqnarray}
Note that these expression use the covariant fields defined in
Eqs.~(\ref{covAbb}-\ref{covpibb}), and thus we write:
\begin{eqnarray} P^{\mu}&=&i\int
d^{4}X~b^{\dag\nu}\partial^{\mu}b_{\nu}\\ M^{\mu\nu}&=&i\int
d^{4}X~\left[b^{\dag\rho}(X^{\mu}\partial^{\nu}-X^{\nu}\partial^{\mu})b_{\rho}+(b^{\dag\mu}b^{\nu}-b^{\dag\nu}b^{\mu})\right]\end{eqnarray}
As in the classical case, the Liouville operator,
$V=n_{\mu}P^{\mu}$, generates translations in the external time
parameter. And it is the external boost generator,
$^{ext}K(m)=n_{\mu}m_{\nu}M^{\mu\nu}$ that is of the most
importance as we can see in its action on foliation dependant
objects: \be
U(\Lambda)\,^nA_{\mu}(X)U(\Lambda)^{-1}=\Lambda^{\nu}_{\mu}\,^{\Lambda
n}A_{\nu}(\Lambda^{-1}X)\ee where $U(\Lambda)=e^{iK(m)}$. The
crucial point here is that it generates Lorentz transformations on
the foliation vector as well. Let us now analyse this issue in a
bit more detail.

Though the set of all coherent states is independent of the
foliation vector, $n_{\mu}$, (they are eigenstates of the
annihilation operator), the definition of them in terms of the
Weyl generator, Eqs.~(\ref{weylApi}-\ref{weylbb}) is clearly not.
It is thus that the Fock space, $\V$, depends upon the choice of
foliation. Now, as in the case of the scalar field, all the
foliation-dependent representations of the history algebra exist
on the same Fock space, $\V$, and $^{ext}K(m)$ relates the objects
defined with respect to a foliation `$n$', with those same objects
defined with respect to the foliation `$\Lambda n$'. For example,
under $^{ext}K(m)$, the constraint operators will transform:
\begin{eqnarray}
^n\Psi_{\kappa}&\underrightarrow{^{ext}K(m)}&\,^{\Lambda
n}\Psi_{\kappa} \\
^n\Phi_{\kappa}&\underrightarrow{^{ext}K(m)}&\,^{\Lambda
n}\Phi_{\kappa}\end{eqnarray}

Now the map $L$ from $\V$ to $\V_{phys}$ is also evidently
$n$-dependent and thus $\V_{phys}$ also depends on the choice of
foliation used to define $\V$. However, whereas all
foliation-dependent representations can exist on $\V$ (and we can
thus talk about transformations between then), the physical
Hilbert spaces, $^n\V_{phys}$ and $^{\Lambda n}\V_{phys}$ (where,
we trust, the point of the added superscript is self-evident), are
clearly different. This is why there will be no representation of
$^{ext}K(m)$ on $\V_{phys}$. Mathematically (and analogously to
the classical case) this situation is represented by the fact that
the external boost does not (weakly) commute with either of the
constraints.

Of course, we can still relate the important quantities on
$^n\V_{phys}$ such as the action, $^nS$, to those same quantities
on $^{\Lambda n}\V_{phys}$ via the prescription given at the end
of Section \ref{vphys}, ie. by mapping back to $\V$, boosting
there, and then mapping to $^{\Lambda n}\V_{phys}$

The other generators will be represented on $\V_{phys}$, we just
make use of Eq.~\ref{bVbVphys} to define them. The most important
of these is the Liouville operator, and this will be defined on
the physical Hilbert space as:\be\label{VVphys} \tilde{V}=i\int
d^{4}X~b^{\perp\dag}_{\nu}\partial^{\mu}b^{\perp\nu}\ee This will
generate time transformations in the external time label, `$t$',
on the physical Hilbert space. We thus arrive, using
Eqs.~\ref{HVphys} and \ref{VVphys}, at the definition of the
action operator on the physical Hilbert space: \be\label{SVphys}
\tilde{S}~=\tilde{V}~-~\tilde{H}\ee

\section{Conclusion}

The aim of this paper has been to construct a histories theory of
Electromagnetism working in the HPO consistent histories
framework. As a vector field theory with two first class
constraints, we have built on the work of Savvidou \cite{sav01a}
on scalar field theory, as well as demonstrating an application of
the constrained systems algorithm developed by Savvidou and
Anastopolous \cite{sav01c}.

Classically, we defined the histories phase space and the two
Poincar\'{e} groups that are a feature of histories field
theories. The constraints partition the constraint surface (and
indeed the whole phase space) into orbits, and by defining
coordinates that are constant on each orbit, we defined the
reduced phase space that carries the physical degrees of freedom
of the theory. We stressed the importance of the foliation
dependence of the phase space (and thus the reduced phase space)
focussing particularly on the action of the external boost
generator which transforms between different foliations.

Quantising within the Dirac scheme, we first constructed the
Hilbert space of the unconstrained theory ($\V$), motivated by
finding a suitable representation of the History algebra. We then
defined the constraints as operators, and, making use of the
technology of coherent states, sought to define the physical
Hilbert space ($\V_{phys}$). As the constraints have continuous
spectra, this was not going to be a true linear subspace of the
full Hilbert space. We got around this issue by analysing $\V$ in
terms of coherent states, which lead to a definition of
$\V_{phys}$ in terms of just the transverse components of the
vector field and their conjugate momenta (or, more strictly, the
space of test functions). We then defined a suitable mapping from
$\V$ to $\V_{phys}$ and used this to define the action operator on
${\V_{phys}}$.

\bigskip
\bigskip

\noindent\large\textbf{ACKNOWLEDGEMENTS} \normalsize

I would chiefly like to thank Ntina Savvidou for her endless
patience and encouragement, but thanks are due also to Chris Isham
and Jonathan Halliwell for helpful discussions.

This work is supported by PPARC PPA/S/S/2002/03440.


\newpage

\begin{thebibliography}{99}

\bibitem{sav01a} K.~Savvidou. Poincar\'{e} Invariance for
Continuous-time Histories. \emph{J.~Math.~Phys} 43:3053, 2002.
gr-qc/0104053.
\bibitem{sav99b} K.~Savvidou. Continuous Time and Consistent
Histories. PhD Thesis, Imperial College, 1999. gr-qc/9912076.
\bibitem{sav01c} K.~Savvidou and C.~Anastopoulos. Histories
Quantisation of Parameterised Systems 1: Development of a General
Algorithm. \emph{Class.~Quant.~Grav} 17:2463, 2000. gr-qc/9912077.
\bibitem{sav01b} K.~Savvidou. General Relativity Histories Theory:
Spacetime Diffeomorphisms and the Dirac Algebra of Constraints.
\emph{Class.~Quant.~Grav} 18:3611, 2001. gr-qc/0104081.
\bibitem{sav03a} K.~Savvidou. General Relativity Histories Theory
1: The spacetime character of the canonical description.
gr-qc/0306034.
\bibitem{sav03b} K.~Savvidou. General Relativity Histories Theory
2: Invariance groups. gr-qc/0306036.
\bibitem{nol01a} D.~Noltingk. Classical History Theory of Vector
Fields. \emph{J.~Math.~Phys} 43:3036, 2002. gr-qc/0107067.
\bibitem{nol01b} D.~Noltingk. BRST Quantisation of Histories
Electrodynamics. gr-qc/0110121.
\bibitem{gri84} R.~B.~Griffiths. Consistent histories and the
Interpretation of Quantum Mechanics. \emph{J.~Stat.~Phys.}
36:219-272, 1984.
\bibitem{omn88} R.~Omn\`{e}s. Logical Reformulation of Quantum
Mechanics 1 Foundations. \emph{J.~Stat.~Phys.} 53:893-932, 1988.
\bibitem{gmh90} M.~Gell-Mann and J.~B.~Hartle. Quantum Mechanics in
the light of Quantum Cosmology. In Complexity, Entropy and the
Physics of Information, ed. W.~Zurek. Addison-Wesley, Reading,
1990.
\bibitem{gmh93} J.~B.~Hartle. Spacetime Quantum Mechanics and the
Quantum Mechanics of Spacetime. Proceedings on the 1992 Les
Houches School, Gravitation and Quantisation, 1993. gr-qc/9304006.
\bibitem{ish94} C.~J.~Isham. Quantum Logic and the Histories Approach to Quantum
Theory. \emph{J.~Math.~Phys.} 35:2157, 1994. gr-qc/9308006.
\bibitem{il94} C.~J.~Isham and N.~Linden. Quantum Temporal Logic
and Decoherence Functionals in the Histories Approach to
Generalised Quantum Theory. \emph{J.~Math.~Phys.} 35:5452-5476,
1994. gr-qc/9405029.
\bibitem{il95} C.~J.~Isham and N.~Linden. Continuous Histories and
the History Group in Generalised Quantum Theory.
\emph{J.~Math.~Phys.} 36:5392-5408, 1995. gr-qc/9503063.
\bibitem{ilss98} C.~J.~Isham, N.~Linden, K.~Savvidou and
S.~Schreckenberg. Continuous Time and Consistent Histories.
\emph{J.~Math.~Phys.} 37:2261, 1998. gr-qc/9711031.
\bibitem{sav99a} K.~Savvidou. The Action Operator for Continuous
Time Histories. \emph{J.~Math.~Phys.} 40:5657, 1999.
gr-qc/9811078.
\bibitem{ara} H.~Araki. Hamiltonian formalism and the canonical
commutation relations in quantum field theory.
\emph{J.~Math.~Phys.} 1:492, 1960.
\bibitem{dir64} P.~A.~M.~Dirac. Lecture on Quantum Mechanics.
Belfer, New York, 1964.
\bibitem{wipf} A.~W.~Wipf. Hamilton's Formalism for Systems with
Constraints. Lectures given at the seminar `The Canonical
Formalism in Classical and Quantum General Relativity', Bad
Honnef, September 1993. hep-th/9312078.
\bibitem{ht92} M.~Henneaux and C.~Teitleboim. Quantization of
Gauge Systems. Princeton University Press, Princeton, 1992.
\bibitem{itz} C.~Itzykson and J-B.~Zuber. Quantum Field
Theory. McGraw-Hill, Singapore, 1980.
\bibitem{ks} J.~R.~Klauder and B.~Skagerstam. Coherent States:
Applications in physics and mathematical physics. World
Scientific, Singapore, 1985.
\bibitem{rs} M.~Reed and B.~Simon. Methods of Modern Mathematical
Physics 1: Functional analysis. Academic Press, 1980.
\end{thebibliography}
\end{document}